# Spotify Danceability and Popularity Analysis using SAP

Authors: Virginia Ochi, Ricardo Estrada, Teezal Gaji, Wendy Gadea, and Emily Duong
FEMBA, California State University Los Angeles
BUS 5100-93 Intro to Business Analytics
vfrazee@calstatela.edu restra62@calstatela.edu tgaji@calstatela.edu wgadea@calstatela.edu nduong@calstatela.edu

**Abstract:** Our analysis reviews and visualizes the audio features and popularity of songs streamed on Spotify*. Our dataset, downloaded from Kaggle and originally sourced from Spotify API, consists of multiple Excel files containing information relevant to our visualization and regression analysis. The exercise seeks to determine the connection between the popularity of the songs and the danceability. Insights to be included and factored as part of our analysis include song energy, valence, BPM, release date, and year.

## 1. Introduction

For the purpose of this project, we used Kaggle Datasets. We were able to find and combine multiple datasets which include raw data with information that is broken down by year, elements of music, genre, artist, and region. The rich information in our files allowed us to tell a story and predict the danceability and popularity. Music streaming was accelerated during the COVID pandemic, with estimates showing a spike of nearly 35% (CounterPoint). We chose this topic because as we all have worked, in some manner, from home and have relied on music and streaming to pass the time. In this paper, our aim is to provide analysis as if we were a band manager partnering with Spotify's data scientists. As band managers, we need to understand trends in music and use this data to forecast where and what type of music we should release.

## 2. Related Work

Spotify launched in 2008 just a year after the first iPhone. The idea was for Spotify to become the soundtrack to a listener's day, offering a range of playlists and news podcasts in the mornings. The algorithm and use of data has very much evolved since then. An artist account has access to data that can help them find an audience, create tailored marketing and touring strategies, and track monthly listeners. The platform also finds new music on behalf of listeners based on their internal data. Spotify Analytics then offers the artist and the label a detailed report about track performance - but to rank, an artist across the board is important to gather and aggregate data from different sources. Soundchart.com for example uses stats and Analytics tools to study performance from different sources. For the outcome of this project, we are using Spotify data to understand the algorithm and how some of the most popular and danceable songs make it to the top of the music charts.

## 3. Specifications

The main file is the Spotify Generic data, it includes data from 1990 to present and detailed information about more than 175,000 songs. The information was collected from Spotify Web API. The dataset includes a unique value (the track ID generated by Spotify), 12 columns with numeric values that measure specific elements of a song for example energy, duration, tempo, loudness, popularity, etc., and categorical data like Artist, key, and release date. The second file is also from Kaggle, but the data was extracted from organizedourmusic.playlistmachinery.com. It contains a list of the top songs by country by Spotify based on the Billboard charts. 13 variables can be explored by looking at this dataset: bpm, energy, danceability, pop, and the country, just to mention a few. Table 1 contains basic information about the files used during our analysis/visualization project.

*Table 1. Data Specification*

| Data Set | Size |
|---|---|
| Spotify Genetic Data | 33.1 MB |
| Top 50 | 118 KB |

## 4. Implementation Flow chart

After an extensive search for quality data, we choose the two CSV files from Kaggle because it includes valuable information that helps analyze the industry. The process of data examination and manipulation is shown in the flowchart (Figure 1). The two different excel files were uploaded to Sap. Once the datasets were available we examined for anomalies, inconsistencies and removed information that was not needed for the purpose of our project. The file with the Genetic data we use for our story charts and line graphs. The second file (Top 50 Geo) we used to create geo-visualization.

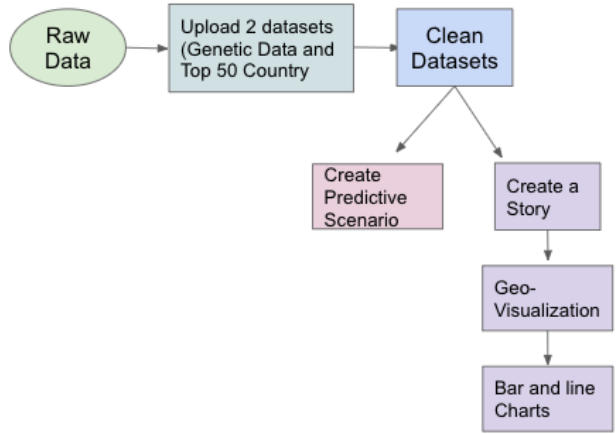

Figure 1. Implementation Flowchart

### 5. Data cleaning

Once the data was uploaded to SAP we realized that some of the information was not needed for the purpose of our project. We Downloaded raw Spotify data from Kaggle which included songs by popularity, danceability, valence, tempo, energy, and more identifiers. The file had 33.1MB of data, so we wanted to throw out some unnecessary factors such as the years we were not using and duplicative information. We also split the DD-MM-YYYY columns into individual day, month, and year columns so we could track trends on an annual basis. We were able to do the clean-up with the tools on SAP.

### 6. Analysis and Visualization
### a. Spotify Elements / Predictive Scenario

In order to analyze the danceability and popularity of the songs, we first had to understand how the data was measured by the app. What we learn is that the collection of the data is broken down by very specific elements of music. For our project, we compared all influencers to a limited number of influencers (5).

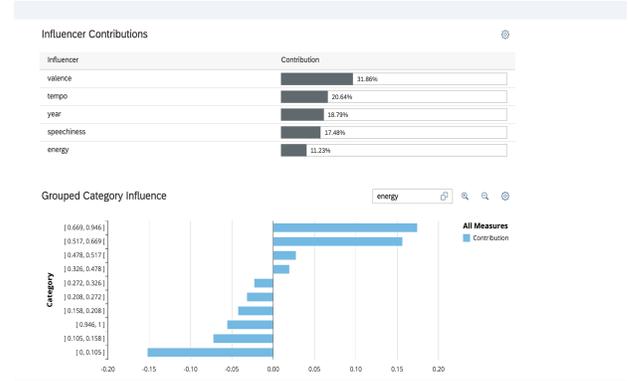

Figure 2. Regression Model 1

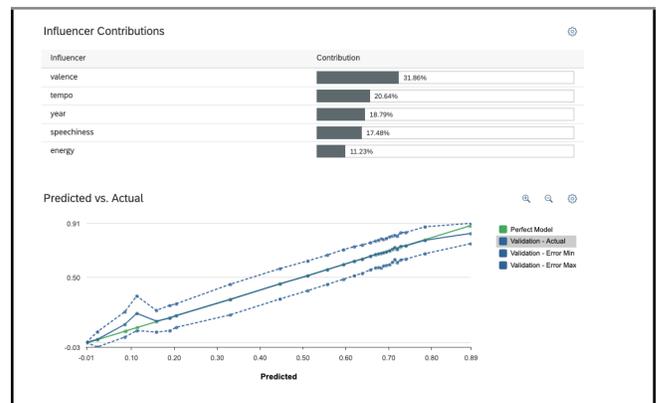

Figure 3. Regression Model 2

After digesting Fig 3. We learned that the top influencers in a regression model for danceability are Valence and Tempo. Thus, concluded that a song with a high measure in those categories has a high probability of hitting top music charts. With the predictable scenario in this model we can predict how danceable a certain year will be, then compare this predictive model to the actual data.

Predictive Model 1: Learned that Valance is the highest contributor (24.93%). Accuracy: 98.17
Predictive Model 2: Valence is still the highest @ 31.86% Accuracy: 97.81

### b. Danceability Predicted by Year

In this story, we created a Comparison (Combination Column and Line) chart between the actual danceability and the Predicted Value. The results were very close (as seen below). We segmented the Dimensions by Release Year.

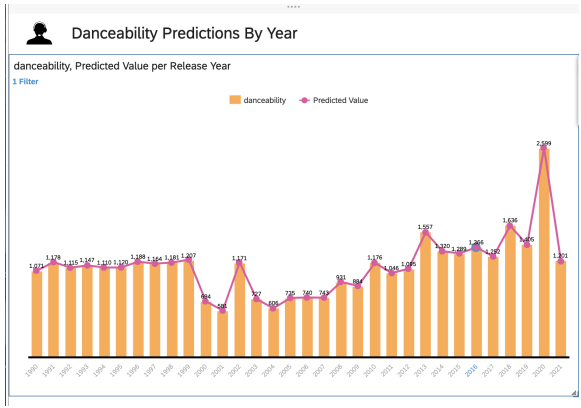

*Figure 4. Danceability Prediction by Year*

### c. Geographic Analysis

Even though music is an international language, the taste, beat, and energy preferred by different countries is not the same. After learning how to use different sets of data in our class Intro to Bus Analytics with Dr. Woo, we added a supplementary data set (Spotify Top 50 Geo) to our existing story. This dataset included the top songs by country for years 1990-present. We had to identify the country column as location data using Actions > Geo Enrichment, doing this allows SAP Analytics to recognize this column as an Area Name. We created a chart titled Top Songs by Geo by inserting a chart - > Geo Map and selected a choropleth/drill layer. We then compared the countries with the most popular songs against Danceability. In the same manner, we added a bubble layer to the Geo Map to compare countries with the songs against most energy. This map revealed that North America had the least danceable and energetic popular songs and South America had the most danceable and energetic popular songs.

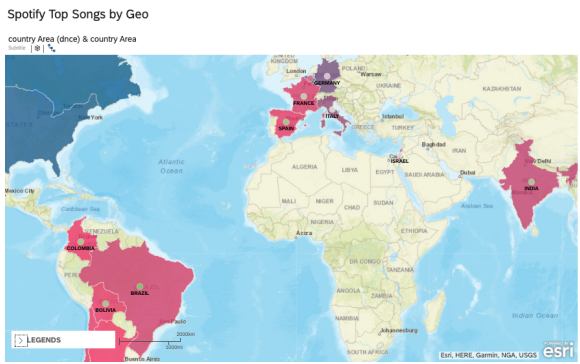

*Figure 5. Spotify Top Songs by Geo*

With the same data, a grouping analysis (Figure 6) shows that by clustering the data based on popularity by BPM and energy, we can see which countries are most likely to respond well to a new hit single. Using the popularity by country data, we created a new responsive story page with a correlation - bubble chart. For the x/y axis, we measured BPM against energy using the popularity as the size dimension.

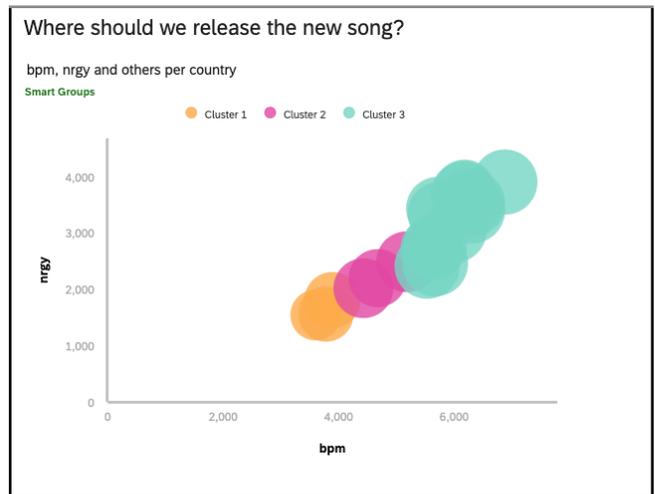

*Figure 6. Cluster popularity by Region*

To create the clusters, we used Smart Grouping and limited it to 3 clusters then cleaned up the data following step d in lab 6

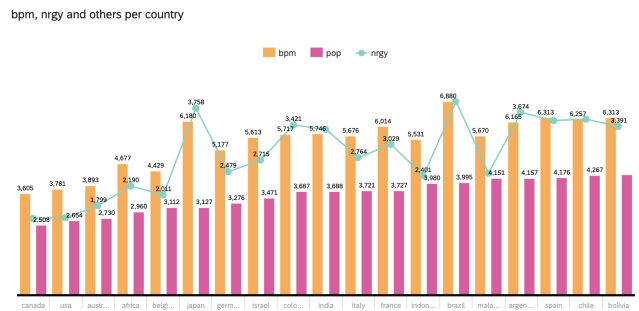

*Figure 7. Cluster popularity by Region*

This revealed that there are 9 countries in the cluster, which groups based on high energy and high bpm (beats per minute) that would be ideal for tours for songs with high energy and bpm.

### d. Historical Timeline/Analysis

Through the analysis, we found that the annual average danceability between 1990 – 2021 saw a moderate increase. We wanted to understand whether danceability would continue to increase. We used time series analysis and forecasted with Triple Exponential Smoothing to find that the average score for danceability will continue to see an increase. Our next time series analysis was to determine whether popular ranked songs will play a factor and become more danceable in the future. We added automatic forecasts to help us determine that the more popular the songs are, the more danceable. We can see that by 2025, the top 10 popular songs will be among the most danceable.

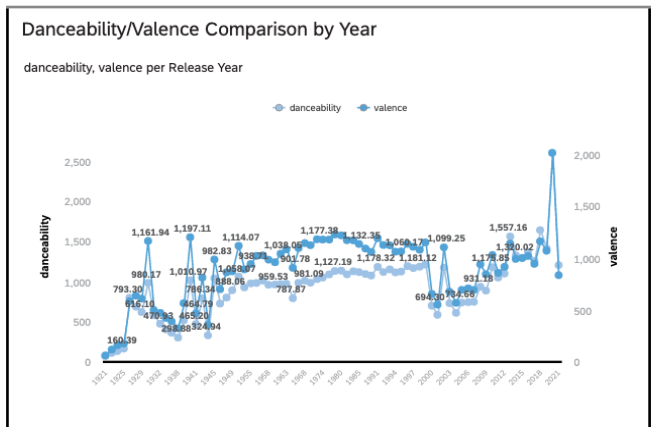

*Figure 8. Danceability Comparison by Year*

### e. Predictability Analysis

We created a times series analysis to review and identify the trends in popularity and danceability. We created two time-series charts leveraging datasets "Spotify by year" and "Average Spotify scores"

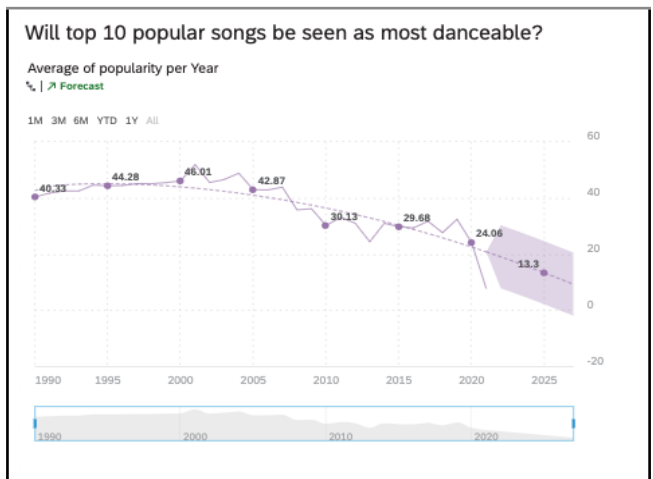

*Figure 9. Top 10 Popular Predictability Model*

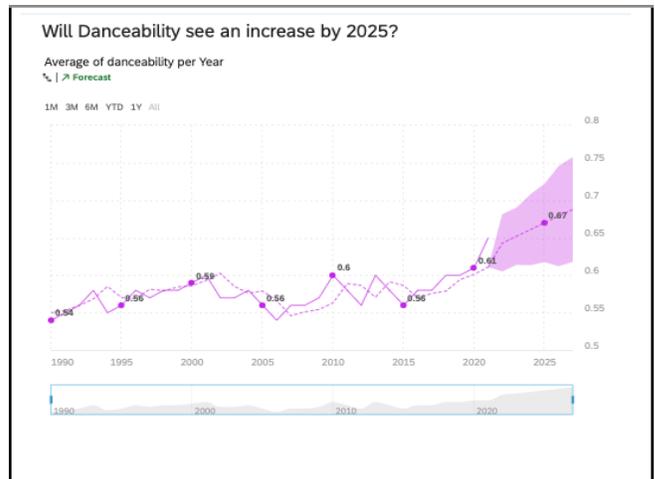

*Figure 10. Danceability Comparison by Year*

Story Location: Cal State LA  folder > Virginiaf > Group 1 Data > Danceability Predictions by Year Story

GitHub: vfrazee/SpotifyDataAnalysis